\documentstyle[11pt]{article}

\def\bichpr{\hoffset=-20truemm
\voffset=-15truemm
\textwidth=16 truecm
\textheight=22 truecm }

\bichpr

\title{3D reduction of the three-fermion Bethe-Salpeter equation }
\author{J. Bijtebier\thanks{{Senior Research Associate at the 
   Fund for Scientific Research (Belgium).}}\thanks{{\it E-mail address:} 
jbijtebi@vub.ac.be}\\Theoretische Natuurkunde, Vrije Universiteit Brussel\\ Pleinlaan 2 B-1050
Brussel, Belgium.}

\begin{document}
\maketitle
\begin{abstract}
We present a 3D approximation of the three-fermion Bethe-Salpeter equation. Our 3D equation is covariantly cluster separable and
the two-fermion cluster separated limits are exact equivalents of the corresponding two-fermion Bethe-Salpeter equations. The
potentials include positive free energy projectors in order to avoid continuum dissolution. 
\end{abstract}

\section{Introduction}
The elimination of the relative times in the three-fermion Bethe-Salpeter equation can be performed in many ways. This equation
can for example be approximated by 3D Schr\"odinger-Pauli or Faddeev equations. In principle a lot of higher-order correction terms
of various origins, often neglected, should restore the equivalence with the initial Bethe-Salpeter equation. We are searching
for a  3D equation which would be an element in a chain of approximations transforming the original Bethe-Salpeter equation
into a manageable equation and would also satisfy at best the following list of requirements: Lorentz invariance, cluster
separability, hermiticity and slow energy dependence of the potentials, correct heavy mass limits, absence of continuum
dissolution. The solutions of the corresponding two-fermion problem will provide the building blocks of our three-fermion
equation.    

\section {The two-fermion problem.} 

The  Bethe-Salpeter equation for the bound states of two fermions is
\begin{equation}\Phi = G_0 K \Phi \label{1}\end{equation}  
where  $\Phi$ is the Bethe-Salpeter  amplitude, $\,K\,$  the Bethe-Salpeter kernel (sum of the irreducible Feynman graphs) and
\begin{equation}G_0=G_{01}G_{02},\qquad G_{0i} = {1 \over p_{i0}-h_i+i\epsilon h_i}\,\beta_i \label{2}\end{equation}  
 the free propagator. The $\,h_i\,$ are the free Dirac hamiltonians
\begin{equation}h_i = \vec \alpha_i\, . \vec p_i + \beta_i\, m_i\qquad (i=1,2).\label{3}\end{equation} 
 We shall denote the total and relative momenta, and the corresponding combinations of the free hamiltonians by
\begin{equation} P = p_1 + p_2\ , \quad p = {1 \over 2} (p_1 - p_2),\quad S = h_1 + h_2\ , \quad s = {1 \over 2} (h_1 -
h_2).\label{4}\end{equation} 
We shall also need the positive and negative free energy projectors:
\begin{equation}\Lambda^{\pm\pm}=\Lambda_1^\pm\Lambda_2^\pm,\quad \Lambda_i^\pm={E_i\pm h_i\over2E_i},\qquad 
E_i=\sqrt{h_i^2}=(\vec p_i^2+m_i^2)^{1\over 2}. \label{5}\end{equation} 
 The free propagator $\,G_0\,$ will be written as the sum of an approached propagator 
$G_\delta\,$ (combining a constraint  fixing the relative energy,  and a global 3D propagator) and a rest $\,G_R.\,$  Salpeter's 3D
propagator, which appears automatically in case of an "instantaneous kernel" is
\begin{equation}\int dp_0 G_0(p_0)\,=\,-2i\pi \,{\Lambda^{++}-\Lambda^{--}\over P_0-S}\,\,\beta_1\beta_2.\label{6}\end{equation}   
 We shall skip the $\,\Lambda^{--}\,$ projector and write the free propagator as 
\begin{equation}G_0=  G_{\delta}+G_R,\qquad G_\delta(p_0)\,=\,-2i\pi \,\delta(p_0\! -\!
s)\,{\Lambda^{++}\over P_0-S}\,\,\beta_1\beta_2.\label{7}\end{equation} 
The Bethe-Salpeter equation  becomes then the inhomogeneous equation
\begin{equation}\Phi=\Psi +G_RK\Phi,\qquad \Psi=G_\delta K\Phi.\label{8}\end{equation} 
 Eliminating $\,\Phi:$
\begin{equation}\Psi=G_\delta K(1-G_RK)^{-1}\Psi=G_\delta K_T\Psi, \qquad K_T=K+KG_RK+...\label{9}\end{equation}   
 The reduction series $\,K_T\,$  re-introduces in fact the
reducible graphs into the Bethe-Salpeter kernel, but with $G_0$ replaced by $G_R$.
The equation becomes
\begin{equation}\Psi=G_\delta K_T\Psi=-2i\pi\,\delta(p_0\! -\! s)\,{\Lambda^{++}\over
P_0-S}\,\beta_1\beta_2\,K_T\,\Psi.\label{10}\end{equation} 
Eliminating the relative energy dependence gives a single 3D equation:
\begin{equation}\Psi=\delta(p_0\! -\! s)\,\psi,\qquad \psi\,=\,{\Lambda^{++}\over P_0-E_1-E_2}\,V(P_0)\,\psi\label{11}\end{equation}
\begin{equation}V(P_0)\,=-2i\pi\,\int dp'_0 dp_0\, \delta(p'_0\!
-\!s)\,\beta_1\beta_2K_T(p_0',p_0,P_0)\,\delta(p_0\! -\!s).\label{12}\end{equation}
We explicitated the dependence of the operator $\,K_T\,$ in the conserved total momentum $\,P_0\,$ and 
in the relative momentum $\,p_0\,$ (this last dependence being non-local). 
\par    Similar results are obtained with  other constraints and 3D propagators.

\section{The three-fermion problem.}
 The Bethe-Salpeter equation is now:
\begin{equation}\Phi=\left[G_{01}G_{02}K_{12}+G_{02}G_{03}K_{23}+G_{03}G_{01}K_{31}+G_{01}G_{02}G_{03}K_{123}\right]
\Phi\label{13}\end{equation} 
where $K_{123}$ is the sum of the purely three-body irreducible contributions. We shall neglect it and
replace the two-body kernels by instantaneous ones, equivalent at the  cluster-separated limits. Among an infinity
of choices, we shall use directly the two-body potentials of the previous section, putting the spectator fermion on the mass shell:
$${K}_{12}(p'_{120},p_{120},P_{120})$$
$$\approx \,\beta_1\beta_2\,\Lambda^{++}_{12}\int dp'_{120} dp_{120}\, \delta(p'_{120}\!
-\!s_{12})\beta_1\beta_2K_{T12}(p'_{120},p_{120},P_0-h_3)\,\delta(p_{120}\! -\!s_{12})\Lambda^{++}_{12}$$
\begin{equation}=\,{-1\over2i\pi}\,\beta_1\beta_2\,\,\Lambda^{++}_{12}\,V_{12}(P_0\!-\!h_3)\,\Lambda^{++}_{12},...\label{14}\end{equation}    
where $\,P\,$ is now the total energy-momentum of the three-fermion system. The Bethe-Salpeter equation becomes
\begin{equation}\Phi={-1\over2i\pi}\,G_{01}G_{02}G_{03}\,\beta_1\beta_2\beta_3\left[\,\Lambda^{++}_{12}\,V_{12}\,\,\Lambda^{++}_{12}\,
\psi_{12}\,+\,\cdots\,+\cdots 
\,\right]\label{15}\end{equation}
\begin{equation}\psi_{ij}(p_{k0})=\beta_k\,G_{0k}^{-1}\int dp_{ij0}\,\Phi.\label{16}\end{equation}
This leads to a set of three coupled integral equations in the $\,\psi_{ij}.\,$  
We shall search for solutions analytical in the Im($p_{k0})\!<\!0\,$ half  planes and close the integration paths
clockwise in these planes. The only singularities will then  be the poles of the free propagators. Performing the
integrations with respect to the $\,p_{ij0}\,$
 gives then
$$\psi_{12}(p_{30})=\,{\Lambda^{++}_{12}\over
(P_0-S)-(p_{30}-h_3)+i\epsilon}\,\left[\,\Lambda^{++}_{12}\,V_{12}\,\Lambda^{++}_{12}\,\psi_{12}(p_{30})\right.$$
\begin{equation}\left. +\,
\Lambda^{++}_{23}\,V_{23}\,\Lambda^{++}_{23}\,\psi_{23}(h_1)\,+\,\Lambda^{++}_{31}\,V_{31}\,\Lambda^{++}_{31}\,
\psi_{31}(h_2)\,\right]\label{17}\end{equation}
and similarly for $\,\psi_{23}\,$ and $\,\psi_{31}.\,$ Solving (\ref{17}) with respect to $\,\psi_{12}(p_{120})\,$ confirms its
analyticity in the Im($p_{k0})\!<\!0\,$ half plane. Furthermore, equation (\ref{17}) shows that the three projections
$\,\Lambda^+_k\psi_{ij}(h_k)\,$ are equal  (let us call them $\,\psi\,$) and satisfy the 3D equation
\begin{equation}\psi={\Lambda^{+++}\over
P_0-E_1-E_2-E_3}\,\left[\,V_{12}(P_0\!-\!E_3)+V_{23}(P_0\!-\!E_1)+V_{31}(P_0\!-\!E_2)\,\right]\,\psi.\label{18}\end{equation}
Moreover, it can be shown that $\,\psi\,$ is the integral of the Bethe-Salpeter amplitude with respect to the relative
times:         
\begin{equation}\psi={-1\over2i\pi}\,\int dp_{10}dp_{20}dp_{30}\,\delta(p_{10}+p_{20}+p_{30}-P_0)\,\Phi.\label{19}\end{equation} 

\section{Conclusions: pro's and con's of our three-cluster equation.}\par
--- The positive-energy projectors included in the equation forbid the mixing of the physical bound states with a continuum
combining positive and negative energy free states (equations suffering of this "continuum dissolution" disease have no
normalisable solutions).\par
--- When the two potentials acting on one of the fermions are "switched off", one gets a free Dirac equation for this fermion and
a correct two-fermion equation for the two other fermions (cluster separability). Furthermore, this last equation is an exact 3D
equivalent of the two-fermion Bethe-Salpeter equation.\par --- We did not specify our reference frame until
now. Our equations are not explicitly covariant, but if we assume that they are written in the three-fermion rest frame, we can
always render them covariant by using the conserved total energy-momentum vector
$\,P.\,$ At the cluster-separated limits, however, the cluster separability requirement forbids the use of this vector. The
equations for the two-fermion clusters must then be covariant. The fact that these equations are exact equivalents of the
covariant two-fermion Bethe-Salpeter equations insures an implicit covariance without introducing
Lorentz boosts by hand.\par
--- The 3D reduction of the two-fermion Bethe-Salpeter equation of section 2 is only an example. The requirement of
preserving the equivalence with the original equation leaves a large freedom which could be used to suit the needs of the
three-fermion phenomenology.\par
--- The potentials are hermitian and their dependence in the total energy is an higher-order effect.\par
--- When the mass of one of the fermions becomes infinite, its presence should
be translated in the equations by a potential (Coulombian in QED) acting on the other fermions. This requirement is only
approximately satisfied. Satisfying it exactly would demand the reintroduction of some of the neglected three-body terms.\par
--- Our two-body potentials are the sum of an infinity of contributions symbolized by Feynman graphs. Keeping only  the first
one (Born approximation) or a finite number of them renders the Lorentz covariance of the two-fermion clusters only approximate. One
can use another 3D reduction based on a covariant second-order two-body propagator of Sazdjian, combined with a covariant substitute
of $\,\Lambda^{++}.$
This leads to a 3D three-cluster equation which is  covariantly Born approximable, but more complicated.\par
--- Our equation can also be written as a set of three Faddeev equations. These Faddeev equations can also be
obtained as an approximation of Gross' spectator model equations. This approximation being of the same order than
these already made in Gross' model, further investigations would be needed to decide which model is closer to the
exact Bethe-Salpeter equation.  
\par --- The higher-order three-body contributions we neglected in our approximation are explicitly given at
the Bethe-Salpeter level. We are presently trying to transform them into correcting terms to our
3D equation. 

\end{document}